\documentclass[aps,10pt,prl,reprint,twocolumn,showpacs,floatfix,superscriptaddress]{revtex4-2}
\usepackage{amsfonts,amsmath,amssymb}

\usepackage[margin=20mm,inner=20mm,marginpar=10mm]{geometry}
\usepackage{graphicx, epstopdf}
\usepackage{microtype}
\usepackage[figuresright]{rotating}
\usepackage{setspace}
\usepackage{textcomp}
\usepackage{times}
\usepackage[normalem]{ulem}
\usepackage[abs]{overpic}
\usepackage{color}

\usepackage[T1]{fontenc}
\usepackage[utf8]{inputenc}
\usepackage{chemformula}
\usepackage{soul}
\usepackage{siunitx}
\usepackage{bm}
\usepackage{booktabs}
\usepackage{multirow}

\newlength{\figwidth}
\begin{document}

\title{Impact of molecular properties on diffraction at nanomasks with low charge density}

\author{Ksenija Simonovi\'c}
\email[]{ksenija.simonovic@univie.ac.at}
\affiliation{University of Vienna, Faculty of Physics, VDS, VCQ, Boltzmanngasse 5, A-1090 Vienna, Austria}

\author{Richard Ferstl}
\affiliation{University of Vienna, Faculty of Physics, VDS, VCQ, Boltzmanngasse 5, A-1090 Vienna, Austria}

\author{Anders Barlow}
\affiliation{University of Melbourne, Faculty of Engineering and Information Technology, Materials Characterisation \& Fabrication Platform, Grattan Street, Parkville, Victoria, 3010, Australia}

\author{Armin Shayeghi}
\affiliation{University of Vienna, Faculty of Physics, VDS, VCQ, Boltzmanngasse 5, A-1090 Vienna, Austria}
\affiliation{Institute for Quantum Optics and Quantum Information (IQOQI), Boltzmanngasse 3, A-1090 Vienna, Austria}

\author{Christian Brand}
\affiliation{German Aerospace Center (DLR), Institute of Quantum Technologies, Wilhelm-Runge-Stra\ss e 10, 89081 Ulm, Germany}

\author{Markus Arndt}
\email[]{markus.arndt@univie.ac.at}
\affiliation{University of Vienna, Faculty of Physics, VDS, VCQ, Boltzmanngasse 5, A-1090 Vienna, Austria}%

\date{\today}

\begin{abstract}   
The quantum wave nature of matter is a cornerstone of modern physics, which has been demonstrated for a wide range of fundamental and composite particles. 
While diffraction at nanomechanical masks is usually regarded to be independent of atomic or molecular internal states, the particles' polarisabilities and dipole moments lead to dispersive interactions with the grating surface.
In prior experiments, such forces largely prevented matter-wave experiments with polar molecules, as they led to dephasing of the matter wave in the presence of randomly distributed charges incorporated into the grating.
Here we show that ion-beam milling using neon facilitates the fabrication of lowly-charged nanomasks in gold-capped silicon nitride membranes.
This allows us to observe the diffraction of polar molecules with a four times larger electric dipole moment than in previous experiments.
This new capability opens a path to the assessment of the structure of polar molecules in matter-wave diffraction experiments.
\end{abstract}

\maketitle

\section{Introduction}
Advanced machining of diffraction masks with nanometer resolution has led to a multitude of optical elements for matter waves.
These include slits, gratings, zone plates and holograms for electrons~\cite{moellenstedtElektronenMehrfachinterferenzenRegelmaessigHergestellten1959,jonssonElectronDiffractionMultiple1974,joenssonElektroneninterferenzenMehrerenKuenstlich1961}, neutrons~\cite{zeilingerSingleDoubleslitDiffraction1988} as well as atoms and dimers~\cite{keithDiffractionAtomsTransmission1988, carnalYoungDoubleslitExperiment1991,shimizuDoubleslitInterferenceUltracold1992, fujitaManipulationAtomicBeam1996, doakRealizationAtomicBroglie1999, luskiVortexBeamsAtoms2021}.
Diffraction at nanomechanical gratings was key for identifying the helium dimer~\cite{schollkopfNondestructiveMassSelection1994} and in the first demonstration of the wave nature of hot \ch{C60} fullerenes~\cite{arndtWaveParticleDuality1999}.
Nanomechanical masks have been successfully employed in full-fledged matter-wave interferometers across a mass range spanning seven orders of magnitude~\cite{KeithInterferometerAtoms1991a,vanderzouwObservationNondispersivityScalar1999,gronnigerElectronDiffractionFreestanding2005, brezgerMatterWaveInterferometerLarge2002, salaFirstDemonstrationAntimatter2019,feinQuantumSuperpositionMolecules2019}.
This wide applicability stems from the fact that they modulate the wavefront of the incident de Broglie wave, and they do this to first order independently of any internal particle property.
Hence, they are considered to be 'universal'. 
\begin{figure}[h]
	\includegraphics[width=0.9 \linewidth]{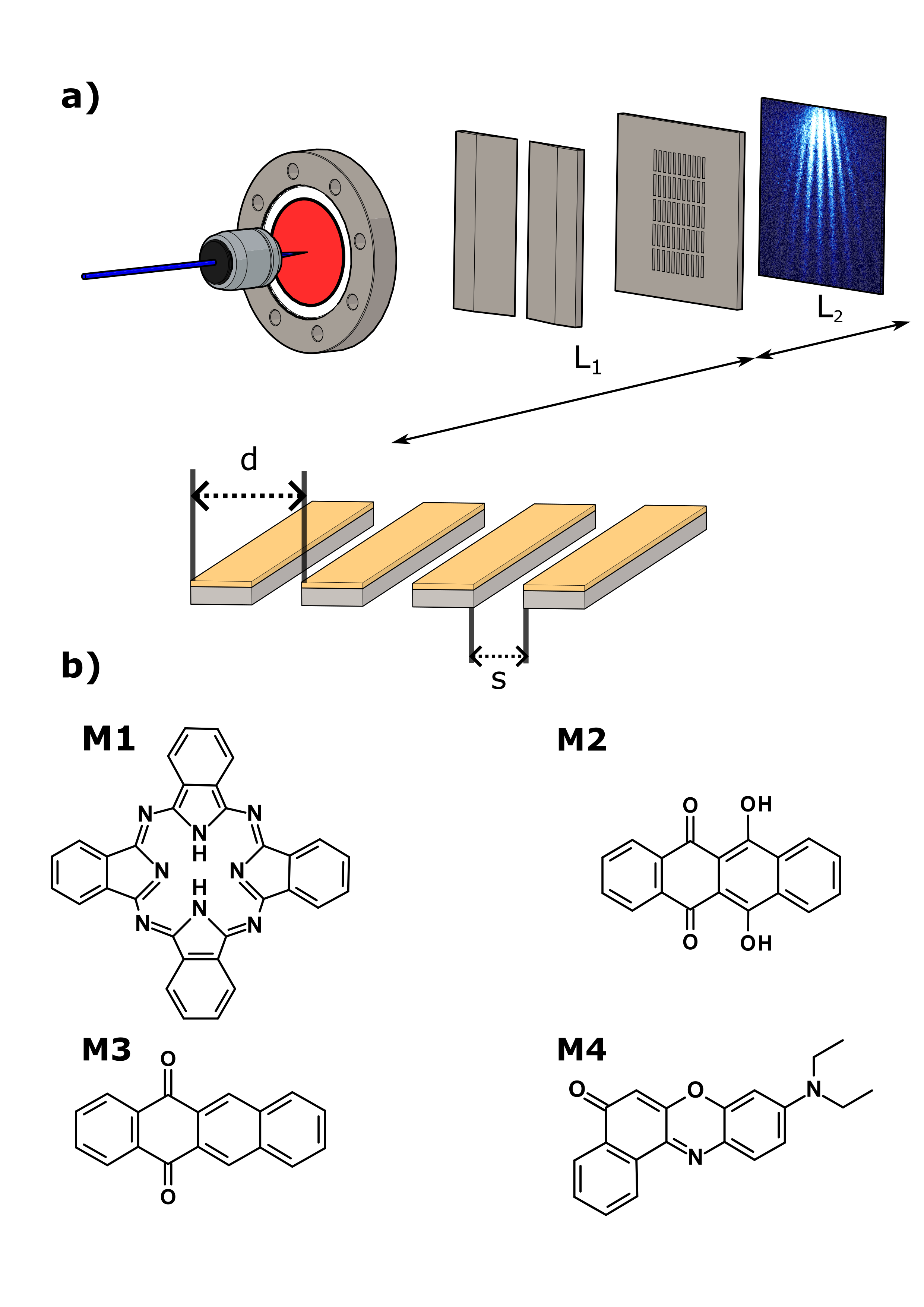}
	\caption{\textbf{a)} Experimental setup for far-field diffraction of polar molecules at a nanomechanical grating. Molecules are sublimated by a micro-focused laser beam and collimated horizontally to sub-\SI{5}{\micro m} before being diffracted at a nanomechanical grating with period $d$ and opening slit width $s$. Molecules diffracted at this mask land on a quartz window and are imaged in fluorescence microscopy. \textbf{b)} Molecules used in this study: phthalocyanine ({\bf M1}), 6,11-dihydroxy-5,12-naphthacenedione ({\bf M2}), 5,12-naphthacenequinone ({\bf M3}), and Nile red ({\bf M4}).}
	\label{fig:Setup}
\end{figure}

However, for applications in electron interferometry nanogratings must be conductive to prevent decoherence due to random potentials caused by trapped charges~\cite{gronnigerElectronDiffractionFreestanding2005, mcmorranDiffraction5keVElectrons2006}. 
Atoms and molecules, on the other hand, are affected by the dispersive Casimir-Polder interaction~\cite{grisentiDeterminationAtomSurfaceVan1999, nairzQuantumInterferenceExperiments2003, garcionIntermediateRangeCasimirPolderInteraction2021}. 
This conservative potential attracts the particles towards the grating walls, thus populating higher diffraction orders than one would expect based on the geometrical slit width. 
For highly polarisable macromolecules, the attraction becomes so pronounced that it may even require substituting mechanical gratings by optical ones~\cite{gerlichKapitzaDiracTalbot2007,nairzDiffractionComplexMolecules2001}. 
To mitigate this effect, gratings have been thinned to the level of single-layer graphene~\cite{brandAtomicallyThinMatterwave2015}.
Nevertheless, even in thin membranes ion beam writing can implant ions or alter the material, which causes additional interactions~\cite{brandGreenFunctionApproach2015, allenGalliumNeonHelium2019a, brandMorphologyDoublyclampedGraphene2021}. 
These might even exceed the Casimir-Polder interaction strength by an order of magnitude, thus dominating the interaction while the particle traverses the grating~\cite{brandGreenFunctionApproach2015}. 
The ensuing phase shift depends on the molecular geometry, velocity, internal degrees of freedom, and in particular on the charge distribution inside the molecule and the membrane~\cite{fiedlerCasimirPolderPotentialsExtended2015,knoblochRoleElectricDipole2017}.
Even though de Broglie diffraction is about the center-of-mass motion of a particle, dephasing associated with random orientations of polar molecules close to nanomechanical masks can suppress the observation of matter-wave interference~\cite{knoblochRoleElectricDipole2017}.

In our present work, we show that we can push the applicability of nanomasks in quantum experiments further than before, specifically to polar molecules with electric dipole moments $p_\mathrm{el}$ beyond \num{8} Debye.
This has been realised by coating insulating silicon nitride membranes with gold and substituting gallium ions with neon in the milling process of the gratings.
While we cannot prevent dephasing entirely, we show that it can be exploited to derive useful information about the molecular structure in the gas phase.
This is exemplified in the analysis of the proton binding sites of 6,11-dihydroxy-5,12-naphthacenedione (6,11-DNHpQ).

\section{Experimental Setup}
Our experimental setup is based on earlier work~\cite{juffmannRealtimeSinglemoleculeImaging2012} and shown in Fig.~\ref{fig:Setup}a).
We coat a thin film of molecules onto the inside of a vacuum window and evaporate them with a laser beam at $\lambda=\SI{421}{nm}$, focused to a spot size below \SI{5}{\micro m}.
The emergent molecular beam is collimated by a vertical mechanical slit to 
 $<\SI{3}{\micro rad}$ divergence angle before it reaches the diffraction grating at $L_1=\SI{0.91}{m}$ behind the source. 
 About $L_2=\SI{0.70}{m}$ behind the grating the diffraction pattern is captured on a $\SI{170}{\micro m}$ thick quartz slide, which closes the high-vacuum chamber. The molecular pattern on the slide is detected using wide-field laser-induced fluorescence microscopy~\cite{juffmannRealtimeSinglemoleculeImaging2012}. 

\begin{figure*}[t]
    \centering
    \includegraphics[width=\linewidth]{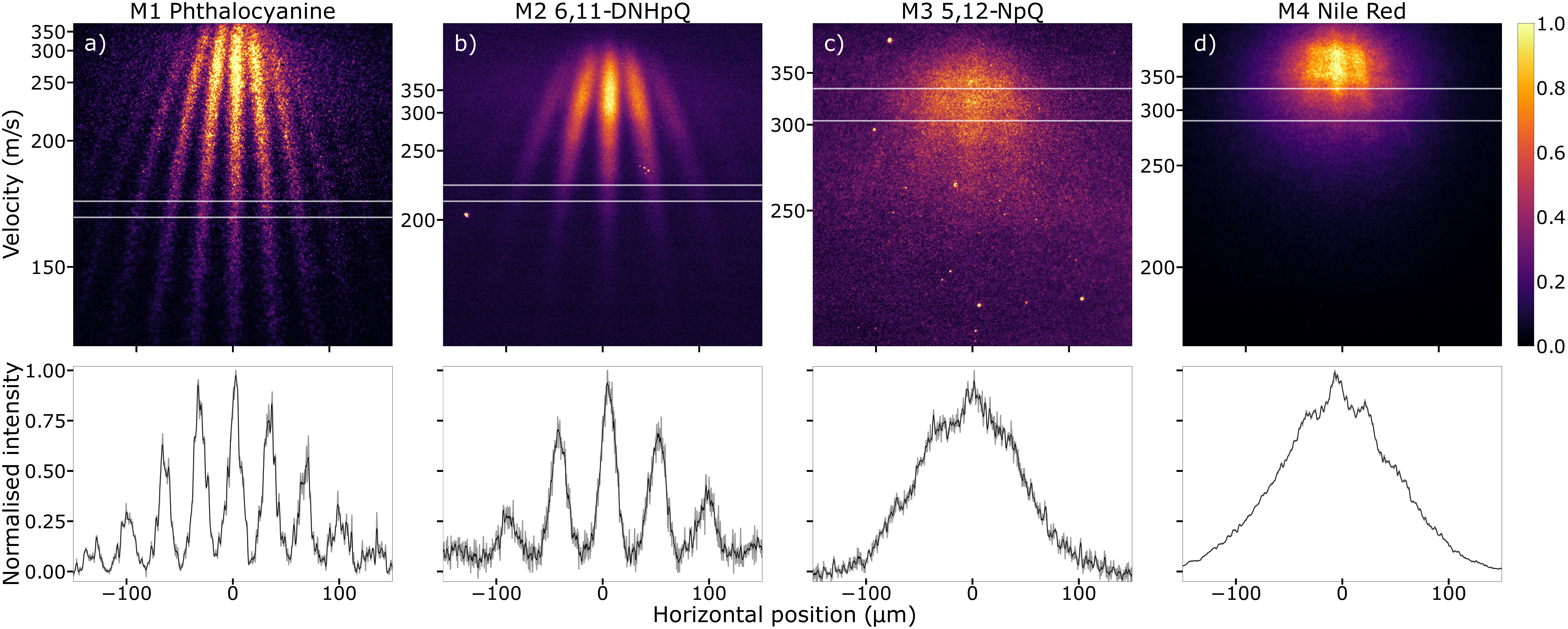}
    \caption{Normalised patterns of the studied systems diffracted at the \SI{20}{nm} thin grating.
    The molecules have been sorted with increasing electric dipole moment $p_\mathrm{el}$ from left to right.
    In the upper row, the diffraction patterns are shown over a detector region of $300\times300 $\si{ \,\micro m^2}.
    The lower row shows exemplary traces that have been vertically binned over the region enclosed by the two white lines in the upper row. The light grey traces in the background are raw data, the black overlays were smoothed using Savitzky-Golay filters for ease of view. The velocity bands were chosen for good contrast and signal strength, centred around $\lambda_\mathrm{dB}=[4.6; 6.3; 5.2; 4.0]\times 10^{-12}\,\rm m$. The velocity scale for M3 and M4 is tentative as the peak separation and signal/noise do not allow for an unequivocal assignment of $v$ over the whole image. All images are background-corrected to account for illumination artefacts.}
    \label{fig:Interferece_images}
\end{figure*}

Previous gratings for such experiments were manufactured either using electron beam writing~\cite{keithDiffractionAtomsTransmission1988}, photolithography~\cite{savasAchromaticInterferometricLithography1995}, or focused gallium ion beam writing (Ga-FIB)~\cite{juffmannRealtimeSinglemoleculeImaging2012}. The present gratings were milled into silicon nitride (SiN$_\text{x}$) using a focused beam of neon ions (Ne-FIB).
This technique is expected to deposit fewer charges in the substrate~\cite{allenGalliumNeonHelium2019a} and thus to be better compatible with polar molecules than gratings written by Ga-FIB.
The thickness of the two employed membranes is \num{15} and \SI{50}{nm} and both were coated with a \SI{5}{nm} thin gold layer on one side to neutralise implanted charges and shield their electric fields.
Both membranes were patterned with a grating period $d$ of about \SI{100}{nm} and a slit width $s$ of \SI{45}{nm}. For further details see Supplementary Information.
  
\subsection{Molecular systems}

To test whether the present gratings are compatible with the diffraction of polar molecules, we study four different systems: phthalocyanine ({\bf M1}), 6,11-dihydroxy-5,12-naphthacenedione ({\bf M2}, abbreviated 6,11-DNHpQ), 5,12-naphthacenequinone ({\bf M3}, abbreviated 5,12-NpQ), and Nile red ({\bf M4}), as shown in Fig.~\ref{fig:Setup}b). They differ largely in their electric dipole moments, from $p_\mathrm{el}=0-8$~Debye, as compiled in Table~\ref{tab:molprop}.  While all molecules were evaporated by the same \SI{421}{nm} laser beam, the fluorescence excitation wavelengths and imaging filters were chosen to match the respective absorption and emission spectra (see Supplementary Information).

To compute the ground state geometries and the respective dipole moments of {\bf M2}, we performed density functional theory (DFT) calculations within the Gaussian 16 program package. We used the LC-$\omega$PBEh functional~\cite{rohrdanzLongrangecorrectedDensityFunctional2009} together with the def2tzvpp basis set~\cite{weigendAccurateCoulombfittingBasis2006, weigendBalancedBasisSets2005}.

\begin{table}[b]
\begin{tabular}{cccr}
\toprule
Label   &  Formula & Mass $(\si{u})$ & Dipole moment $p_{\rm el}$ $(\si{D})$ \\ 
\midrule
\textbf{M1} & \ch{C32H18N8}      & 514.54 & 0.0~\cite{yangGeometricStructureElectronic2022}      \\ 
\midrule
\textbf{M2} & \ch{C18H10O4}      & 290.27 & 0.4 (see Fig.~\ref{fig:ProtonHopping})\\
\midrule
\textbf{M3} & \ch{C18H10O2}      & 258.27 & 0.9 (calc.) / 2.3 (exp.)~\cite{gleicherCalculationsQuinonoidCompounds1974} \\
\midrule
\textbf{M4} & \ch{C20H18N2O2}    & 318.38 & 8.2~\cite{goliniFurtherSolvatochromicThermochromic1998}\\ 
\bottomrule
\end{tabular}
\caption{Chemical composition, mass, and electric dipole moment of the molecules used in this study. For details see text.}
\label{tab:molprop}
\end{table}

\section{Results and Discussion}

Diffracting the non-polar molecule {\bf M1} at the \SI{20}{nm} thin grating leads to well-resolved diffraction peaks up to the fifth order, as shown in Fig.~\ref{fig:Interferece_images}a).
Interestingly, we observe only a slightly stronger population of higher diffraction orders for the 55~nm thick grating (Supplementary Information).
To derive a first estimate for the interaction strength, we fit the patterns with a reduced effective slit width $s_{\text{eff}}$, as this results in a qualitatively similar population of the high diffraction orders~\cite{grisentiDeterminationBondLength2000}.
For the \SI{55}{nm} thick grating, the slit width is reduced from the geometrical $s=\SI{43}{nm}$ to effective $s_{\text{eff}}=\SI{20}{nm}$, corresponding to a reduction factor $s/s_{\text{eff}} = 2.2$.
This is considerably less than the factor of \num{3.3} observed for {\bf M1} diffracted at a \SI{45}{nm} thick grating SiN$_{\text x}$ milled using Ga-FIB~\cite{brandAtomicallyThinMatterwave2015}.
Hence, even though the grating in this study is slightly thicker, the interaction strength is reduced compared to earlier studies.
This is consistent with the assumption that Ne-FIB in combination with gold coating results in fewer implanted charges than Ga-FIB used on uncoated SiN$_\text{x}$ would do.

The question now is whether this allows for coherent diffraction of highly polar molecules, such as {\bf M4} with $p_{\rm el}=\SI{8.2}{D}$.
In earlier experiments, the interference fringe contrast was entirely lost for molecules with a dipole moment of only $p_\mathrm{el}=1.8$~D when diffracting them at a grating milled into an insulating SiO$_2$ membrane using Ga-FIB~\cite{knoblochRoleElectricDipole2017}.
Furthermore, the SiO$_2$ membrane in those experiments had a thickness of only \SI{8}{nm} and a grating geometrical slit width of $ s=\SI{82}{nm}$.
Hence, the molecules were on average further away from the membrane and interacted with it for a shorter time.
Nevertheless, sending {\bf M4} onto the \SI{20}{nm} thick grating produced via Ne-FIB (Fig.~\ref{fig:Interferece_images}d), we observe finite interference contrast, despite it having a more than four times higher electric dipole moment than molecules in earlier studies~\cite{knoblochRoleElectricDipole2017}.
In addition, here we observe that increasing the thickness of the grating from \SI{20} to \SI{55}{nm} leads to similar results, suggesting that the amount of implanted charges in our fabrication process is small.  

To qualitatively describe the dephasing interaction, we measure the width of the individual diffraction peaks~\cite{knoblochRoleElectricDipole2017}. 
For non-polar molecules they correspond to the width of the collimated beam. 
Polar molecules, however, show a beam broadening that increases with their dipole moment, the surface charge density, and the transit time through the grating.
For {\bf M1} and {\bf M2}, the peak width is independent of velocity and even drops slightly with $v$ for {\bf M2} (see Supplementary Information).
In the case of {\bf M3} and  {\bf M4}, however, the observed velocity range is too small to make any statements in this respect.
For these systems, we estimate that the width of the collimated beam increases from $15$ to over \SI{30}{\micro m} following diffraction (see Supplementary Information).

\subsection{Impact of molecular geometry on diffraction pattern}

Although dephasing due to local charges is considerably smaller in our present gratings compared to previous experiments~\cite{knoblochRoleElectricDipole2017}, it is not entirely suppressed. 
As the loss of contrast is a function of $p_{\rm el}$, this suggests that we can extract qualitative information about the dipole moment from the observed fringe contrast.
In the following, we exploit this to explore the structure of two test molecules, {\bf M2} and {\bf M3}.

\begin{figure}[t]
    \centering
    \includegraphics[width=\linewidth]{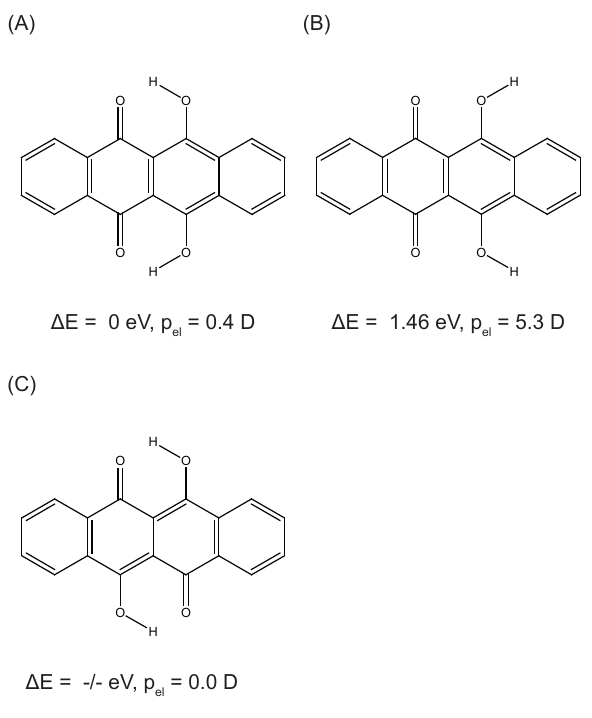}
    \caption{Conformation isomers of 6,11-dihydroxy-5,12-naphthacenedione along with their relative energy $\Delta$E and their dipole moment $p_{\rm el}$.}
    \label{fig:ProtonHopping}
\end{figure}

The two lowest-energy structures of 6,11-dihydroxy-5,12-naphthacenedione ({\bf M2}), obtained from our DFT calculations, are shown in Fig.~\ref{fig:ProtonHopping}.
In structure (A) the OH groups are rotated toward the carbonyl groups, leading to an overall dipole moment of \SI{0.4}{D}.
In contrast to this, when both OH-bonds are pointing away from the neighbouring carbonyl groups, as shown in structure (B), the dipole moment increases by more than an order of magnitude to \SI{5.3}{D}. 
The corresponding diffraction pattern (Fig.~\ref{fig:Interferece_images}b) exhibits high interference fringe contrast, which is only compatible with a very small dipole moment, thus excluding structure (B).
Determining the width of the diffraction orders, we observe no velocity-dependent broadening, but a slight narrowing of the peak width with decreasing $v$ (Supplementary Information). 
This is surprising, as the interaction of randomly oriented polar molecules with electric fields always leads to peak broadening~\cite{knoblochRoleElectricDipole2017}.
One possible explanation is that {\bf M2} is present in structure (C), which exhibits an inversion centre and thus has no dipole moment. 
For the closely related molecule naphthazarin ({\bf M2} without the two outer hexagon rings) this conformational isomer is a stable minimum, lying \SI{270}{meV} higher in energy than (A)~\cite{jezierskaNonCovalentForcesNaphthazarin2021a}.
However, even at a temperature of \SI{800}{\degreeCelsius} in the source, the population of this conformational isomer is on the order of a few percent.
Hence, it is unlikely that it is responsible for the observed pattern.
Another possibility is that the protons hop between the two binding sites on a timescale faster than the transit time through the grating, averaging the dipole moment out.
In an aqueous solution, proton hopping is known to occur within 2~ps~\cite{lightFundamentalConductivityResistivity2004, meiboomNuclearMagneticResonance2004, yuanTrackingAqueousProton2019}.
For {\bf M2} traveling at \SI{300}{m \per s} through a \SI{20}{nm} thick grating the transit time is \SI{70}{ps}, which is long compared to the hopping time.
Additionally, the hopping may be facilitated by the thermal sublimation in the source.
While we cannot study the dynamics of the process in our setup, this explanation does not stand in contrast to our observations.

For {\bf M3}, one may expect the overall dipole moment to be close to zero, as the partial dipole moments of the two carbonyl groups seem to compensate each other.
However, as shown in Fig.~\ref{fig:Interferece_images}c) the observed matter-wave dephasing is much stronger than for {\bf M2}.
While earlier theoretical (\SI{0.87}{D}) and experimental works (\SI{2.3}{D}) differed~\cite{gleicherCalculationsQuinonoidCompounds1974} we observe a strong matter-wave dephasing, which points towards a dipole moment of at least \SI{2}{D}.
This corroborates the prior experimental value. 
%\textcolor{red}{This needs the results from Armin's calculations.}

In all these experiments, one may speculate about the influence of molecular temperature.
Thermally induced electric dipole moments have been observed in earlier matter-wave experiments with functionalized diazobenzenes, where they had a strong effect~\cite{gringInfluenceConformationalMolecular2010}. For phthalocyanine ({\bf M1}) the observation of high-contrast matter-wave fringes indicates that it is too stiff to induce sizeable thermal dipole moments. 
A similar geometric stability is predicted for the aromatic ring systems of {\bf M2} and {\bf M3}. 
In Nile red, the thermal contribution is expected to be non-negligible.  

In earlier matter-wave interferometry experiments, measurements of atomic polarizabilites~\cite{holmgrenAbsoluteRatioMeasurements2010, lonijAtomDiffractionReveals2010} and molecular magnetism~\cite{feinNanoscaleMagnetismProbed2022a} were calibrated with specifics atoms. 
Hence, it is tempting to propose the same for diffraction at the current gratings employing diatomic molecules of well-known dipole moments as a reference.
However, besides the magnitude of the dipole moment $p_{\rm el}$, the interaction depends on partially unknown parameters, such as excited rotations and vibrations, which hampers an exact determination of $p_{\rm el}$.
Nevertheless, it allows us to estimate its magnitude, giving valuable insights into molecular structures in the gas phase.
While electric deflectometry in combination with laser gratings circumvents the issue of an unknown charge distribution inside the grating, it faces the same challenges regarding the molecular parameters~\cite{eibenbergerElectricMomentsMolecule2011a}. 
Moreover, for the molecules we investigate here, the comparatively small optical polarizabilities necessitate highly intense lasers to ensure an effective grating. 

\section{Conclusions}
Writing nanomechanical gratings using focused neon ion beams and coating the dielectric with a thin layer of gold has allowed us to observe interference of highly polar molecules at nanomechanical gratings.
Such systems were so far inaccessible to matter-wave diffraction at mechanical masks. 
While dispersive forces close to the grating wall are often perceived as an obstacle to quantum experiments, we exploit this to shed light on the structure of the molecules, here via their electric dipole moment.
Single grating diffraction is thus an easy and surprisingly powerful tool to study aspects of molecular structure.
Interestingly, this information is accessible even though quantum delocalisation prevents us from knowing the exact location of the molecule in real space.
Comparing interference patterns of the same molecule behind different gratings should also provide information about surface charges implanted during the writing process or accumulated during the deposition process. 
Matter-wave diffraction of polar molecules is thus a sensitive tool to study such effects. 

 Mechanical nanogratings will remain important for the manipulation of atoms and small molecules of low polarisability and low ionisation yield, where optical phase or depletion gratings can only be realised by lasers of high power or low wavelength. Neon-FIB in combination with gold-coated silicon nitride seems to implant fewer charges, resulting in smaller surface forces. Because of that, gratings with periods even below \SI{50}{nm} seem to be in range. Such a tiny period cannot be achieved by optical diffraction gratings, and it would be relevant for the realisation of 2D diffraction masks for lightweight, fast, and polar molecules. 

% *****************************
\section{Acknowledgements}
We acknowledge financial support by the Austrian Science Funds (FWF) through project DOC.85 (HiDHyS) as well as by the FWF project 32542-N. The work was in part conducted at the Materials Characterisation and Fabrication Platform (MCFP) at the University of Melbourne and the Victorian Node of the Australian National Fabrication Facility (ANFF).  
\newpage
%\section{References}
%\bibliographystyle{apsrev4-1}
%\bibliography{dipolesbib}

%

\end{document}

\title{Supplementary information for\\Impact of molecular properties on diffraction at nanomasks with low charge density}
\maketitle

\section{Imaging the diffraction patterns}

\noindent
The diffraction pattern on the quartz window is visualised by exciting the molecules with resonant laser radiation. 
The fluorescence light is collected with a 20x magnification objective, separated from the scattered excitation laser with suitable optical filters and recorded using an electron-multiplying CCD camera. 
All relevant parameters for the investigated systems are compiled in Tab.~\ref{tab:suppl_imaging}.

\begin{table}[ht]
\begin{tabular}{@{}ccc@{}}
\toprule
\textbf{Molecule} & \textbf{\begin{tabular}[c]{@{}c@{}}Excitation\\ wavelength (nm)\end{tabular}} & \textbf{\begin{tabular}[c]{@{}c@{}}Fluorescence\\ filter used (nm)\end{tabular}} \\ \midrule
\textbf{M1}       & 661                                                                           & 711/25                                                                           \\
\textbf{M2}       & 421                                                                           & 550/88                                                                           \\
\textbf{M3}       & 421                                                                           & 550/88                                                                           \\
\textbf{M4}       & 532                                                                           & >550                                                                             \\ \bottomrule
\end{tabular}
\caption{Imaging parameters for molecules used in the study: excitation laser and the fluorescence range observed.}
\label{tab:suppl_imaging}
\end{table}

\section{Manufacturing of the gratings}

\noindent
The nanogratings were milled into two silicon nitride (SiN$_\text{x}$) membranes with a thickness of \SI{15}{nm} and \SI{50}{nm} respectively. 
In both cases the membrane was coated with a \SI{5}{nm} thin film of gold before it was processed.
The writing was done using a \SI{25}{keV} neon ion beam on an \textsc{Orion NanoFab} (Zeiss, Germany), with the gratings design and patterning control performed by the \textsc{NPVE} software package (Fibics Inc., Canada). 
Each grating covers an area of \SI{10}{\micro m} (width) times \SI{20}{\micro m} (height) into which the slit array is written.
These dimensions also define an effective horizontal and vertical collimation delimiter for the molecular beam.
The period of the gratings amounts to $d\approx$\SI{100}{nm} and the geometrical slit opening $s$ is $\approx$\SI{45}{nm}.
The structure is stabilised by $w_h = \SI{0.6}{\micro m}$ wide horizontal support bars. 

\begin{figure}[ht]
\centering
\includegraphics[width=0.5\linewidth]{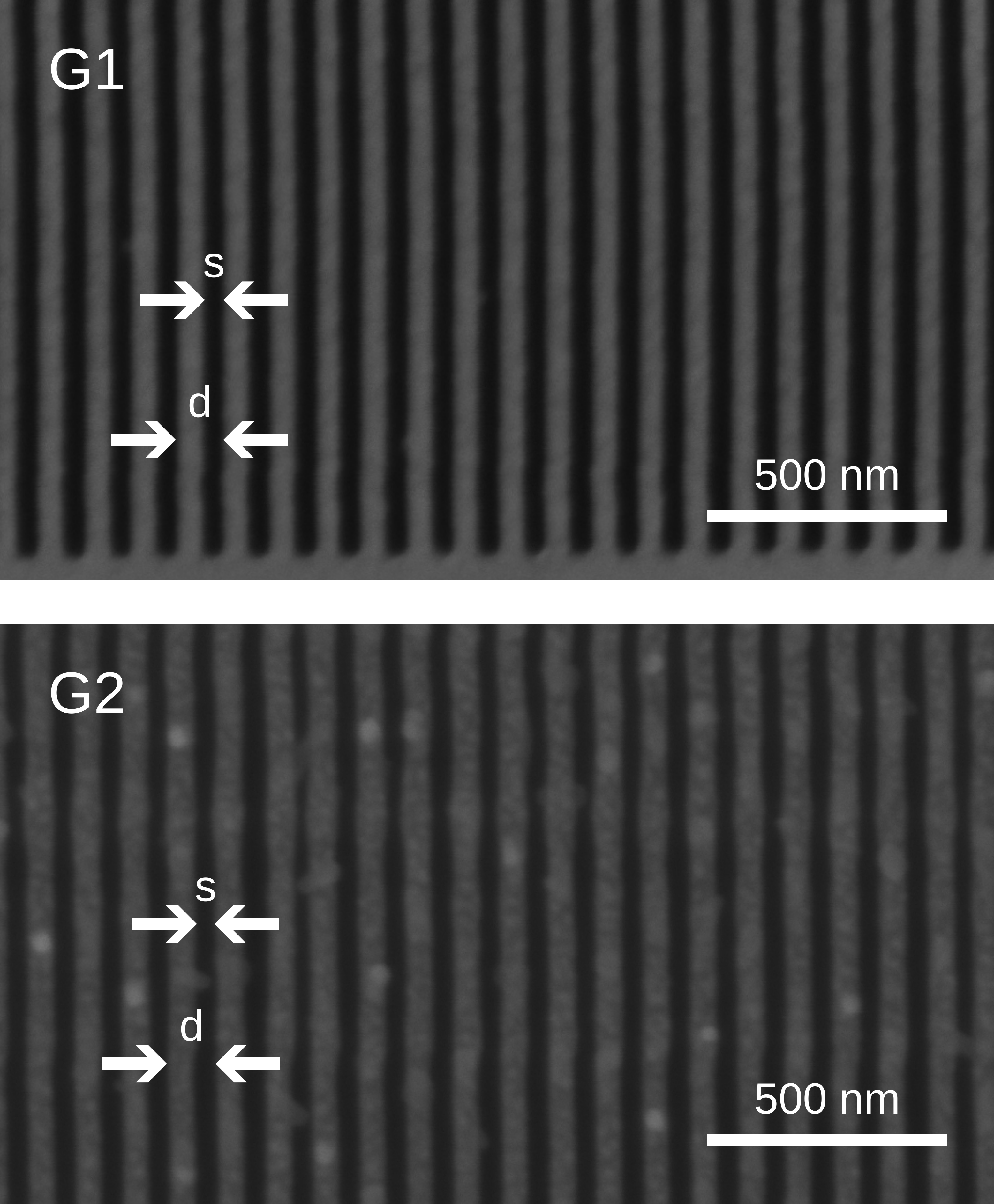}
	\caption{Scanning electron images of gratings \textbf{G1} (\SI{20}{nm} thickness, top) and \textbf{G2} (\SI{55}{nm} thickness, bottom). The parameters $d$ and $s$ denote the grating periodicity and the geometric slit width, respectively.}
	\label{fig:grating-sem-new}
\end{figure}

To ascertain the grating parameters, the diffraction masks were imaged using using scanning electron microscopy (\ref{fig:grating-sem-new}).
From these images we derive the average period and slit width, as compiled in Tab.~\ref{tab:grating-properties}. 
To do this, the images are converted to greyscale, vertically integrated, and normalised.
The minima in the resulting trace correspond to the openings in the grating. 
A periodically spaced array of Gaussian functions is fitted to these dips. 
The period extracted from the fit corresponds to the grating period $d$. 
The slit width $s$ represents the full width at half maximum (FWHM) of the Gaussian peaks.
Local contamination may influence the effective slit width and the envelope of the diffraction pattern, but leave the fringe structure unaffected.

\begin{table}[]
\begin{tabular}{@{}cccc@{}}
\toprule
Grating     & Thickness (\unit{nm}) & Period $d$ (\unit{nm}) & Slit width $s$ (\unit{nm}) \\ \midrule
\textbf{G1} & \num{20}            & \num{97\pm2}         & \num{46\pm2}             \\
\textbf{G2} & \num{55}            & \num{99\pm2}         & \num{43\pm2}             \\ \bottomrule
\end{tabular}
\caption{The geometrical properties for grating G1 and G2 are extracted from SEM images by fitting to traces of 11 slits. The uncertainties are limited by the resolution of the SEM images.}
\label{tab:grating-properties}
\end{table}

\section{Diffraction at 55~nm thick grating}

\begin{figure*}[t]
    \centering
    \includegraphics[width=\linewidth]{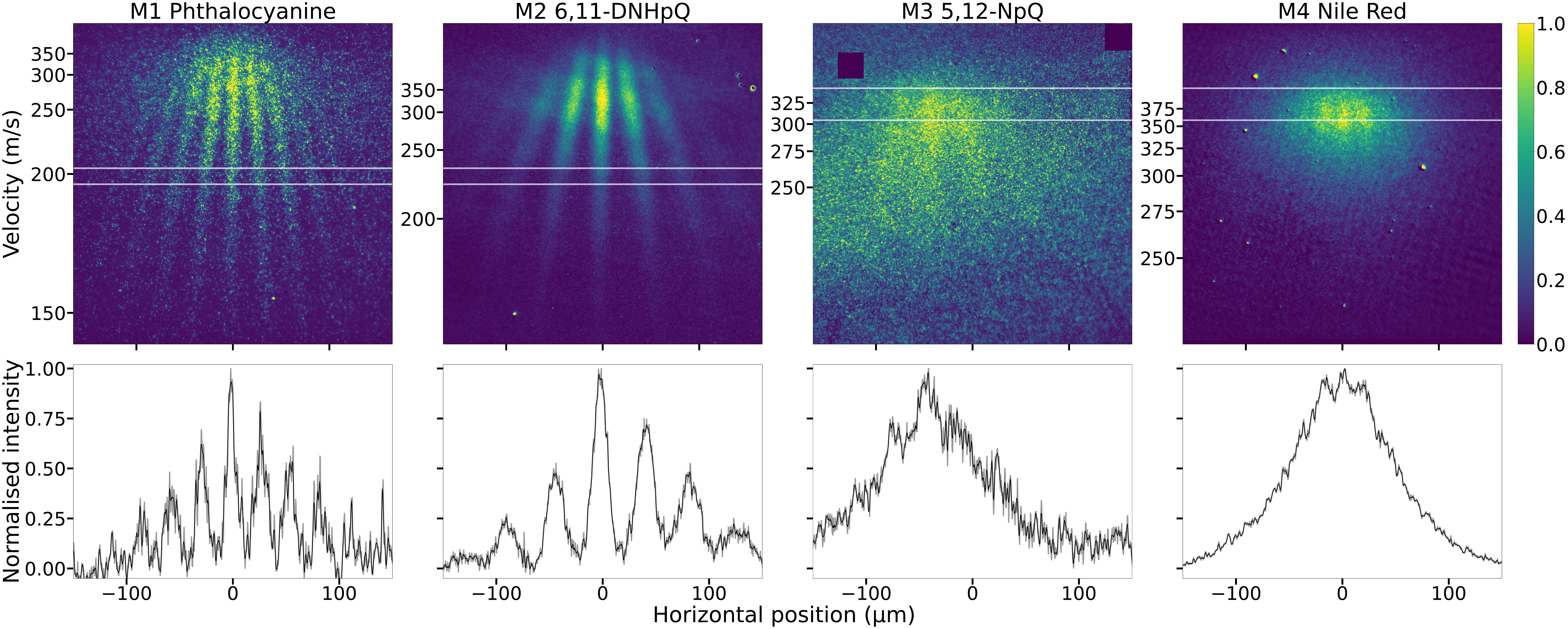}
    \caption{Normalised patterns of the studied systems diffracted at the \SI{55}{nm} thin grating. The molecules are sorted in order of increasing permanent electric dipole moment. Lower panels contain vertically integrated traces of regions marked with white lines. The velocity scale for \textbf{M3} and \textbf{M4} is tentative as the peak separation and signal/noise do not allow for an unequivocal assignment of $v$ over the whole image. In the lower panel the light grey traces in the background are raw data, the black overlays were smoothed using Savitzky-Golay filters for ease of view.}
    \label{fig:suppl_interference-50nm}
\end{figure*}

\noindent
All molecules used in the study were diffracted both at at the \SI{55}{nm} and the \SI{20}{nm} thick grating. 
The respective diffraction patterns obtained from the \SI{55}{nm} thick grating are shown in Fig.~\ref{fig:suppl_interference-50nm}.
Furthermore, they contain vertically binned traces illustrating the width of the diffraction orders.

\section{Fitting the effective slit width for {\bf M1}}

\noindent
We extract the effective slit width $s_{\rm eff}$ for diffraction of phthalocyanine through the \SI{55}{nm} thick grating by fitting the trace with an ideal grating diffraction model. 
\begin{equation*}
    I(x) = c_1 \cdot \sinc^2 \left( \frac{\pi s_{\text{eff}}}{\lambda_{\text{dB}}}\cdot\frac{x}{z}\right) \cdot \left( \frac{\sin \left(N \cdot \frac{\pi d}{\lambda_{\text{dB}}}\cdot \frac{x}{z} \right)}{\sin \left( \frac{\pi d}{\lambda_{\text{dB}}} \cdot \frac{x}{z} \right)}\right)^2 
    + c_2
\label{eq:suppl_fit-model}
\end{equation*}
where $I(x)$ is the normalised intensity distribution with constants $c_1, c_2$, $\sinc(y)=\sin(y)/y$, $\lambda_{\text{dB}}$ the molecules' de Broglie wavelength, $N$ the number of coherently illuminated slits, $d$ the grating period, and $z$ the grating-detector distance. 
For the velocity band \SIrange{210}{230}{m \per s} this results in $s_{\rm eff}\approx \SI{20}{nm}$, showing a considerably smaller reduction compared to previous experiments on similar gratings produced using Ga-FIB~\cite{brandAtomicallyThinMatterwave2015}.
\begin{figure}[ht]
    \centering
    \includegraphics[width=\linewidth]{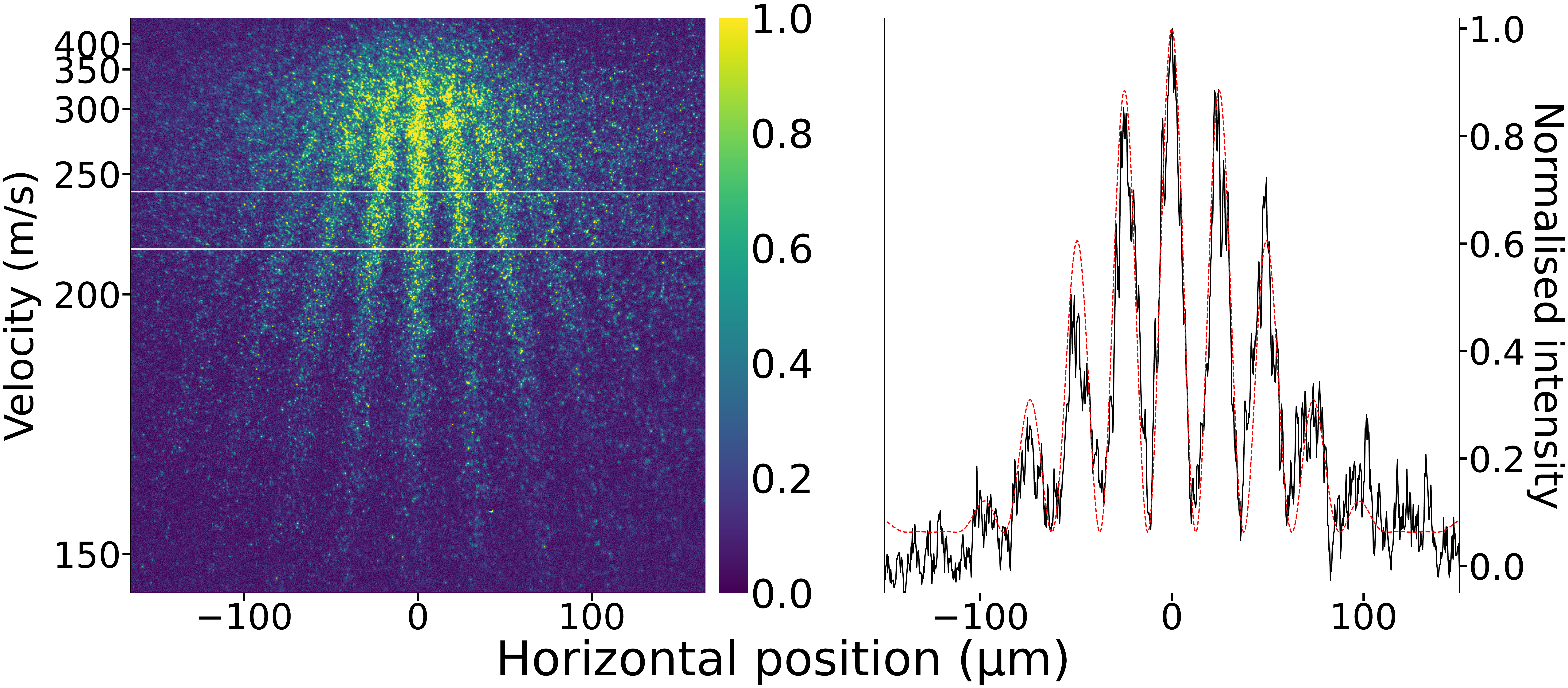}
    \caption{Diffraction image of {\bf M1} phthalocyanine on \SI{55}{nm} thick grating and a normalised vertically integrated trace for the \SIrange{210}{230}{m \per s} velocity band. The effective slit width as extracted from an ideal grating diffraction model is about \SI{20}{nm}.}
    \label{fig:suppl_phthalo-slit-width}
\end{figure}

\section{Broadening of diffraction peaks}

\begin{figure}[h]
\centering
\includegraphics[width=\linewidth]{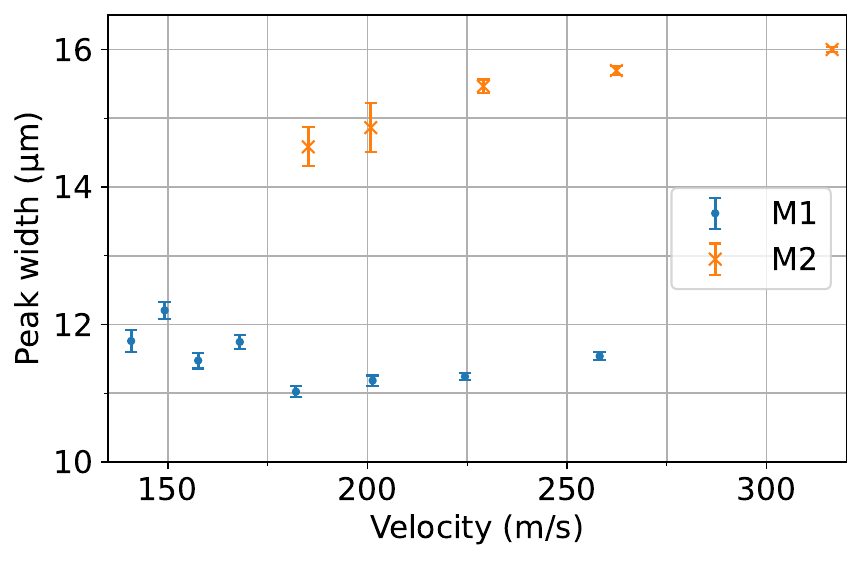}
	\caption{Velocity dependence of the peak widths for the molecules \textbf{M1} and \textbf{M2}. The broadening is practically independent of the velocity for the low dipole moments in these cases. The widths were extracted by fitting equally spaced Gaussians to traces of the diffraction patterns. The corresponding velocities were computed from the spacing between the maxima.}
	\label{fig:peak_width_velocity}
\end{figure}

The diffraction of polar molecules through charged gratings leads to dephasing of the matter wave as each molecule is deflected depending on its orientation and rotational state~\cite{knoblochRoleElectricDipole2017}.
The more time the molecule spends close the the grating walls, the larger the deflection is. 
Hence, the peak broadening increases with decreasing velocity.
To quantify this, we extract the widths $w$ of the diffraction peaks by fitting a sum of equally spaced Gaussians of the form
\begin{equation*}
     y(x) = y_0 + \sum_{i=-N}^N A_i\cdot \exp\left\{-2\frac{(x-i\cdot d)^2}{w^2}\right\}
\end{equation*}
to vertical traces of the diffraction patterns.
%and compare $w$ to the width of the collimated molecular beam not interacting with the grating. 
%This allows us to estimate the molecular dipole moment $p_{\text{el}}$ as a function of peak broadening: stronger interaction between $p_{\text{el}}$ and the grating should lead to more significant peak broadening, implying a higher value of $p_{\text{el}}$. 
For {\bf M1} diffracted at the \SI{20}{nm} thin grating we observe no velocity-dependent broadening, corroborating its lack of dipole moment. 
Also for {\bf M2}, the width of the peaks is virtually independent of the velocities of the molecules as can be seen in Fig.~\ref{fig:peak_width_velocity}. 
Here we even see a decrease in peak width with decreasing velocity. 
%This further corroborates the small electrical dipole moments of the molecules, as we would expect the broadening of the peaks to scale with the interaction time for strongly polar molecules.

\begin{figure}[h]
    \centering
    \includegraphics[width=\linewidth]{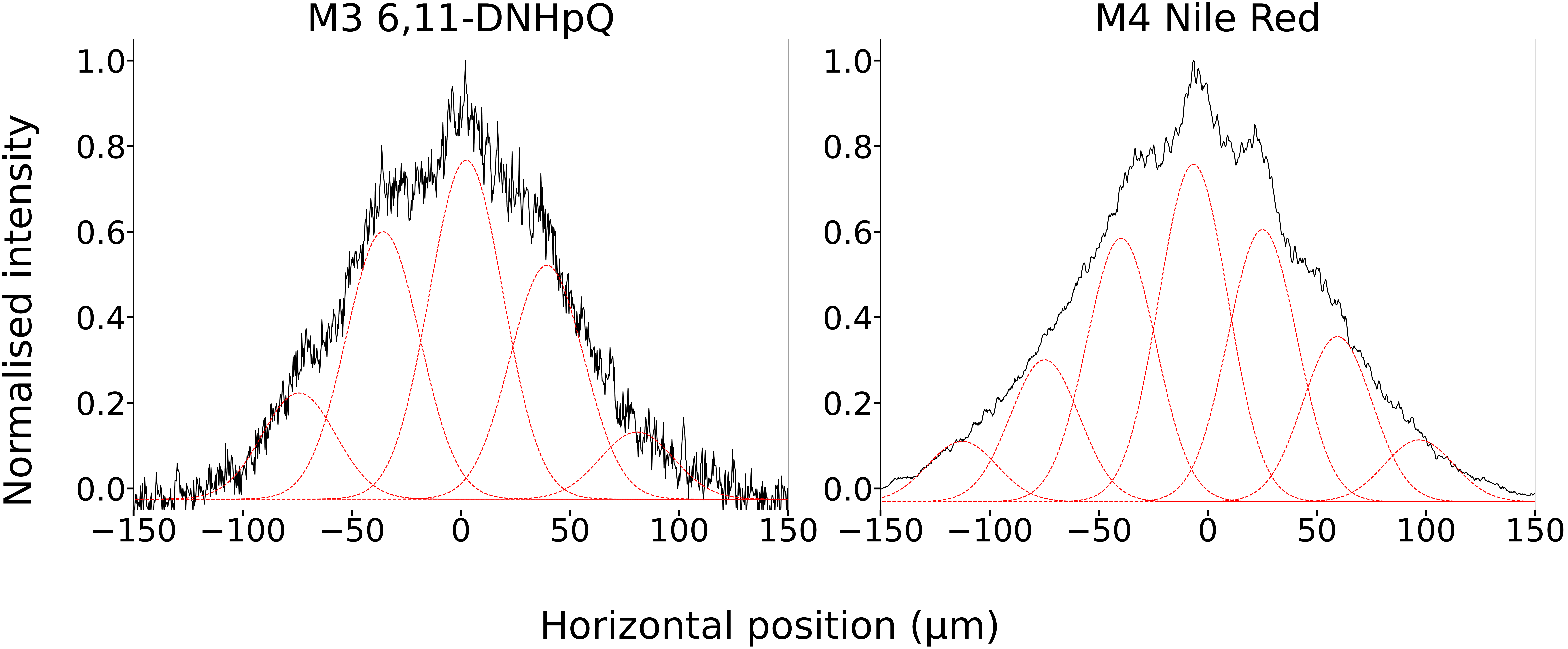}
    \caption{Fitting sums of Gaussians to diffraction patterns of polar molecules \textbf{M3} and \textbf{M4} at the \SI{20}{nm} thick grating to extract the diffracted peak widths. Summed up traces of molecules with velocity bands centred about \SI{300}{m\per s} are used to estimate the strength of the interaction of the molecules' permanent dipole moment with the grating, and hence give information about the magnitude of the dipole moment.}
    \label{fig:suppl_gaussian-widths}
\end{figure}

In the case of {\bf M3} and {\bf M4}, the diffraction pattern does not span a sufficiently large velocity range to extract the velocity-dependence of $w$.
Hence, we compare the width of the undiffracted beam to the one after diffraction.
For both molecules, the beam doubles in width from about \SI{15}{\micro m} to more than \SI{30}{\micro m}. 
The respective fits are shown in Fig.~\ref{fig:suppl_gaussian-widths}. 

%\section{References}